\documentclass[%
 reprint,
 amsmath,amssymb,
 aps,
prl,
floatfix,
]{revtex4-2}

\usepackage{graphicx}
\usepackage{dcolumn}
\usepackage{bm}

\begin{document}


\title{New source for tuning the effective Rabi frequency discovered in multiphoton ionization}

\author{Wankai Li, Yue Lei, Xing Li, Tao Yang, Mei Du, Ying Jiang, Jialong Li, Aihua Liu, Lanhai He, Pan Ma}
\affiliation{Institute of Atomic and Molecular Physics, Jilin University, 2699 Qianjin Avenue, Changchun City, 130012, China
}%
\author{Sizuo Luo}
\affiliation{Institute of Atomic and Molecular Physics, Jilin University, 2699 Qianjin Avenue, Changchun City, 130012, China
}%
\author{Dongdong Zhang}%
 \email{dongdongzhang@jlu.edu.cn}
\affiliation{Institute of Atomic and Molecular Physics, Jilin University, 2699 Qianjin Avenue, Changchun City, 130012, China
}%


\author{Dajun Ding}
\email{dajund@jlu.edu.cn}
\affiliation{Institute of Atomic and Molecular Physics, Jilin University, 2699 Qianjin Avenue, Changchun City, 130012, China
}%


\date{\today}

\begin{abstract}
The Autler-Townes effect due to near resonance transition between 4s-4p states in potassium atoms is mapped out in the photo-electron-momentum distribution and manifests itself as a splitting in the photo-electron kinetic energy spectra. The energy splitting fits well with the calculated Rabi frequency at low laser intensities and shows clear deviation at laser intensities above 1.5$\times$10$^{11}$~W/cm$^{2}$. An effective Rabi frequency formulae including the ionization process explains the observed results. Our results reveal the possibility to tune the effective coupling strength with the cost of the number of level-populations. 
\end{abstract}

\maketitle


One of the great challenges of modern science is to develop tools to manipulate the quantum states of a physical system. Precise quantum state control heavily relies on the accurate knowledge of the coupling strength between the system and the external driving sources. Arguably, variety of laser sources have become the working horse for the this purpose, e.g.,
in physical chemistry, shaped laser pulses are designed to direct chemical reactions along a desired pathway\cite{brif_control_2010}.

On the other hand, the interaction of atoms and laser field has been studied over 40 years, and the resulting photoelectron spectra attract much interest both theoretically and experimentally. Over the last decades, this interest has grown significantly, which benefit from the rapid progress of laser technology, namely, the possibility of generating extremely short and intense pulses~\cite{krausz09a}. The interaction between strong laser fields and atoms or molecules have brought us a lot of surprises, e.g., the above-threshold ionization (ATI)~\cite{agostini79a}, high-harmonic generation (HHG)~\cite{winterfeldt08a}, coherent state manipulation~\cite{vitanov01a}. Rabi-oscillation is one of such phenomena, which is of fundamental interest to modern quantum physics and considered one of the most manifestations of coherent light-matter interactions. Its particularity is related to the ability to form the basis for many applications such as atomic clocks~\cite{ludlow15a}, quantum computing and information processing~\cite{singer10a,derevianko11a,zwanenburg13a,georgescu14a,altman21a}, and quantum control at the electronic level~\cite{kral07a,vitanov17a}. The oscillation frequency named as Rabi frequency is a measure of the coupling strength between two levels in external fields. In the dressed picture the Rabi frequency responds for the light shift and the energy splitting (Autler-Townes (AT) effect) of the coupled levels~\cite{autler_stark_1955}. 

On single photon resonance, the AT effect features the linear scaling law with respect to the amplitude of the driving fields and the repulsive nature of the splitted energy levels~\cite{cohen-tannoudji_nobel_1998}. 
The AT effect in atomic and molecular ionization have been studied in strong pulsed laser fields, e.g., dynamic control of the interference of AT splitting has been observed for potassium atoms interacted with strong time-delayed femtosecond laser pulse trains~\cite{wollenhaupt03a}, the AT splitting in the multiphoton resonance ionization spectrum with ultrashort laser pulses has also been proposed in molecules~\cite{sun_autler-townes_2003} and has recently been observed in polyatomic molecules~\cite{kim_experimental_2020}. 

In this letter, a detailed account for the Autler-Townes (AT) effect measured in the photoelectron momentum distribution of multiphoton ionization of potassium atoms is reported. Beyond the normal AT splitting, at higher intensities we observe significant deviation from the Rabi frequency. An effective Rabi frequency formulae taking into account the ionization process is employed to reproduce the experimental observations, which indicates that decay processes causing loss of populations will affect the Rabi frequency which in turn change the coupling strength.

\begin{figure}[htpb]
\includegraphics[width=0.5\textwidth]{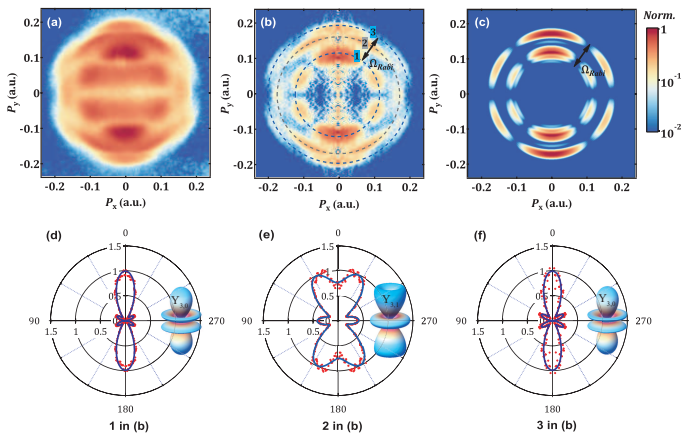}    
\caption{Photoelectron momentum (a-c) and angular (d-f) distribution. 
(a) and (b) show the original electron momentum distribution and the Abel-inverted image. The rings labeled in (b) indicate three separated momentum components. (c) is 
numerical results by solving the time-dependent Schr\"odinger equations (TDSE). The size of the black arrows in (b) and (c) represents the magnitude of the Rabi frequency $\Omega_{Rabi}$. (d)-(f) give the angular distributions for the three rings labeled in (b) corresponding to ring 1-3, respectively. The angular distribution of spherical harmonics $Y_{3,0}(\theta, \phi)$ and $Y_{3,1}(\theta, \phi)$ are shown alongside.}
\label{fig:imaging}
\end{figure}

The detailed experimental setup used in our research has been described in previous papers~\cite{liu19a,Hu20a}. Briefly, the potassium atoms generated from an effusive beam source are irradiated by an 800~nm femtosecond laser with a pulse duration of 50~fs (FWHM). Photoelectrons produced via the multiphoton ionization are detected with a Velocity-Map-Imaging (VMI) apparatus~\cite{chandler87,eppink97}. To confine the laser-atom interaction volume, a thin slit (0.5~mm) is mounted before the atoms fly into the detection region. A typical photoelectron momentum distribution at laser intensity of 1.0$\times$10$^{11}$~W/cm$^{2}$ is shown in Fig.~\ref{fig:imaging}. In Fig.~\ref{fig:imaging}, (a) and (b) are the original and inverted momentum distribution within the momentum range of $\pm$0.25~a.u., and (c) is the calculated results by solving the time-dependent Schr\"odinger equations (TDSE). 
Fig.~\ref{fig:imaging} (d)-(f) show the angular distributions extracted from the momentum distribution for the ring 1-3 labeled in (b), respectively. 3D angular distributions for spherical harmonics $Y_{3,0}(\theta, \phi)$ and $Y_{3,1}(\theta, \phi)$ are plotted alongside for comparison.
The ionization potential of a Potassium atom is 4.34~eV. To elaborate the 4s electron absorption of at least three 1.55~eV photons (800~nm) is required. The lowest order photoelectron should then have its kinetic energy at round 0.3~eV (0.15~a.u.). However, in our experiment three distinct rings in the electron momentum distribution were observed. 
The angular patterns of ring 1-3 (Fig.~\ref{fig:imaging} (d)-(f)) can be grouped into two classes. The ring 1 and the ring 3 exhibit the same angular distribution mimicking a spherical harmonic $Y_{3,0}(\theta, \phi)$, where as the ring 2 shows a $Y_{3,1}(\theta, \phi)$ distribution. For three-photon ionization of a 4s electron in a linear polarized laser field, dipole selection rule indicates that the final photoelectron angular distribution should have the character of spherical harmonics $Y_{3,0}(\theta, \phi)$ and $Y_{1,0}(\theta, \phi)$. Further more the Fano propensity rule suggests that the ionization path favors the increasing of orbital quantum number $l$~\cite{cohen_interference_1966,busto_fanos_2019}. Thus, for three-photo ionization, $Y_{3,0}(\theta, \phi)$ will dominate over $Y_{1,0}(\theta, \phi)$ as shown in Fig.~\ref{fig:imaging}(d) and (f). Similar angular distributions for the ring 1 and 3 indicate that their source are of the same symmetry. Multiphoton ionization of potassium atoms at ultrashort pulsed laser with centre frequency near resonance with D1 and D2 transition have been extensively studied previously~\cite{wollenhaupt_interferences_2002,wollenhaupt_femtosecond_2005,wollenhaupt_femtosecond_2006}, and is still an interesting topic~\cite{hockett_complete_2014,bayer_time-resolved_2019,Li_2021}. From previous studies, we understand that the reason for the appearance of the additional peaks at different kinetic energy but similar angular distribution is due to the Autler-Townes effect~\cite{autler_stark_1955} which states that at near resonance condition, the energy levels of the coupled two states will each split into two. The spacing between the separated levels equals to the Rabi frequency
\begin{equation}
\Omega_{Rabi} = \frac{e\langle a|r|b\rangle E_{0}}{\hbar}  
\label{eq:rabifrequency1}
\end{equation}
 where $|a\rangle$ and $|b\rangle$ are the two levels coupled by the oscillating electric field with the amplitude of $E_{0}$, $e$ is the elementary charge and $\hbar$ is the reduced Planck constant. This energy spacing is then mapped out in the photoelectron momentum image as two rings with different radius but the same angular distribution.
 To further check this, we compared our experimental results to numerical non-relativistic TDSE calculations. The time-dependent Schr\"odinger equation is solved within the single-active-electron approximation using a model potential~\cite{masnou-seeuws_model_1982} which reproduces accurately the binding energies and 4s-4p energy difference of the K atom. In this paper we focus on the ring 1 and ring 3 in Fig.~\ref{fig:imaging} (b) and their kinetic energy changes with laser intensities. Fig.~\ref{fig:imaging} (c) shows the calculation results and its excellent agreement with the experiment. 
 The ring in between with $Y_{3,\pm 1}$ character is due to the spin-orbit coupling in the 4p state which mixes magnetic quantum numbers with $m_{l}=0$ and $m_{l}=\pm 1$~\cite{bayer_time-resolved_2019}. 
 One notices that in the numerical calculation the ring 2 in the Fig.~\ref{fig:imaging}
 (b) is not reproduced. Our calculation is made in the non-relativistic frame which omits the spin-orbit effect in the 4p state. This simplification will not effect the analysis of the present study which focus on the ionization channels with $\Delta m_{l} = 0$. However a more detailed study of the generation of ring 2 will be presented in a forthcoming paper.
\begin{figure}[htpb]
\includegraphics[width=0.3\textwidth]{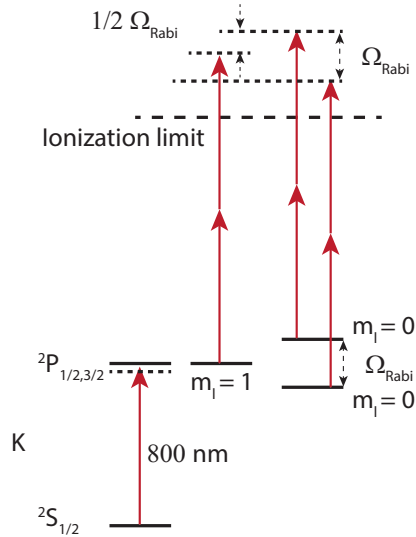}    
\caption{Illustration of the principle of the near resonance multiphoton ionization process. It shows the relevant energy levels and the excitation and the ionization path ways. One 800 nm photon near resonantly couples potassium atom $^{2}S_{1/2}$ and $^{2}P_{1/2,3/2}$ states and results the energy splitting of each level into two. Elaborating an electron from potassium atom in the ground state $^{2}S_{1/2}$ needs three photons which give the free electron wavefunction a $f$ angular distribution character. 
}
\label{fig:principle}
\end{figure}
\begin{figure}[htpb]
\includegraphics[width=0.45\textwidth]{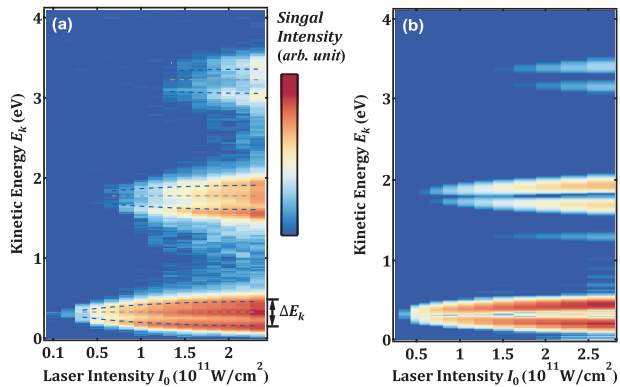}  
\caption{Photoelectron kinetic energy distribution at different laser peak intensities. (a) is the experimental results. (b) is the TDSE calculation results. Tow figures share the same colorbar but with different meanings. The colorbar in (a) indicates the normalized signal intensity. Whereas in (b) it gives the normalized ionization probablities.}
\label{fig:kin}
\end{figure} 


It is well known that coupling of two states with resonance light gives raise the energy splitting of each level as shown in Fig.~\ref{fig:principle}. This effect can be mapped out in photoelectron momentum distribution measurements (Fig.~\ref{fig:imaging}). This is the Autler-Townes effect which was predicted early in 1955~\cite{autler_stark_1955} and subsequently verified in numerous systems~\cite{walker_observation_1995,qi_autler-townes_1999,ahmed_quantum_2011,kim_experimental_2020} and have found its applications in quantum information~\cite{li_dynamical_2012} and information storage~\cite{saglamyurek_coherent_2018}. The energy separation can be expressed as the Rabi frequency 
\begin{equation}
    \Delta E = \Omega_{Rabi} = E_{0} \mu_{0}
    \label{eq:rabifrequency}
\end{equation}
where $E_{0}$ is electric field amplitude of the laser field and $\mu_{0}$ is the transition dipole moment ($\mu_{0} = \langle a|r|b\rangle$ in Eq.~\ref{eq:rabifrequency1}). The energy splitting scales linearly with $E_{0}$ and has been confirmed with photoelectron measurement of ionization of atoms and molecules~\cite{walker_observation_1995,qi_autler-townes_1999,ahmed_quantum_2011,kim_experimental_2020}. In the present paper, we extend the laser intensity and find, for the first time to our knowledge, the significant deviation from this linear scaling law.

\begin{figure}[htpb]
\includegraphics[width=0.3\textwidth]{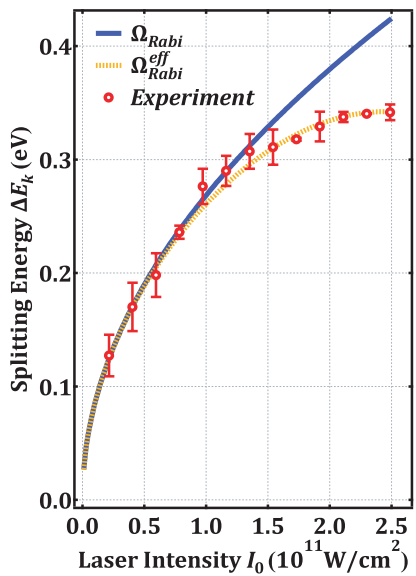}    
\caption{The energy separation of the Aulter-Townes splitting as a function of laser intensities. The red dots represent the experimentally measured kinetic energy difference of the photoelectron originated from the AT-doublets. The error bar is the statistical error from ten individual measurements. The blue line is the Rabi frequency in Eq.~\ref{eq:rabifrequency}. The orange broken line is the effective Rabi frequency in Eq.~\ref{eq:Rabieff}. The excellent agreement between experimental results and the simple theory including the effect of the ionization is clear.}
\label{fig:split}
\end{figure}

In a resonance multiphoton ionization process, once the energy splitting induced by the coupling of the two resonance states exceeds the natural line width $\gamma$ and the spectrum line width $\Delta \omega$ of the laser, the AT-splitting will present in the photoelectron kinetic energy distribution as two separated peaks~\cite{walker_observation_1995,sun_autler-townes_2003} distanted by $\Omega_{Rabi}$. In our experiment, these conditions are fulfilled. At typical laser intensity of $1.2\times10^{11}$~W/cm$^{2}$ in our experiment the Rabi frequency is around 0.3~eV which is observed in Fig.~\ref{fig:imaging} (b). As increasing the laser intensity, the kinetic energies of the two separated peaks go into different directions, i.e., the lower one's kinetic decreases whereas the upper one's increases as shown in Fig.~\ref{fig:kin} (a). This trends reflects the repulsive nature of the splitted levels coupled by a near resonance light~\cite{cohen-tannoudji_nobel_1998}. It is seen that not only the lowest ordered photoelectrons' kinetic energy are splitted but also the their ATIs. Fig.~\ref{fig:kin} (b) shows the TDSE calculation results which reproduce well the experimental results. Again, the spin-orbit effects are neglected in the calculation and we don't expect to obtain the ring 2 laying in between.
To further investigate the laser intensity dependency of the AT-splitting, we extract the value of the kinetic energies by fitting the angular integrated momentum distribution in Fig.~\ref{fig:imaging} (b) with Gaussian profiles at each intensity. The results are shown in Fig.~\ref{fig:split} as the difference of the kinetic energy between the upper and lower components. As clearly observed, at laser intensities below $1.5\times10^{11}$ W/cm$^{2}$ the energy splitting varies as the square root of the intensity which is expected as the Rabi frequency depending linearly on the electric field~\cite{walker_observation_1995,qi_autler-townes_1999,ahmed_quantum_2011,kim_experimental_2020}. As further increase the laser intensity a clear discrepancy between the energy splitting and the Rabi frequency is found. This discrepancy has been predicted theoretically but so far not yet confirmed by experiments~\cite{kaiser_photoionization_2013}.

The Rabi oscillation between two coupled states in an open system will include an additional dumping term caused by the ionization or optical pumping into other states. This dumping term, which is proportional to the ionization or optical pumping rate, will affect the effective coupling strength and can be expressed as~\cite{kaiser_photoionization_2013}:
\begin{equation}
    \Omega_{Rabi}^{eff} = \sqrt{\Omega_{Rabi}^{2}+(i\Gamma)^{2}}.
    \label{eq:Rabieff}
\end{equation}
The multiphoton ionization of potassium atoms has been intensively studied both theoretically~\cite{story_landau-zener_1994,zhang_frequency-modulated_2003} and experimentally~\cite{nicole_saturation_1999,wollenhaupt_interferences_2002,wollenhaupt_femtosecond_2006,hockett_complete_2014,hockett_maximum-information_2015,pengel_control_2017,kerbstadt_control_2019}, from which the ionization rate from both 4s and 4p state can be known.
To simplify the analysis, we consider only ionization from the upper 4p state since the ionization from 4s state requires one more photon. The ionization rate $\Gamma$ from the 4p state depends on the laser intensity quadratically and can be expressed as
\begin{equation}
    \Gamma = \sigma I_{0}^{n}
\end{equation}
where $\sigma$ is the multiphoton ionization cross-section, $I_{0}$ is the laser intensity and $n$ is the number of photons absorbed to free the electron. To get the intensity dependency of $\Gamma$, value of the $\sigma$
is borrowed from the Ref.~\cite{bebb_theory_1967,rahman-attia_two-_1986} and the index $n=1.54$ is derived from experimental measurement of the electron yield vs laser intensity. Plug these values into Eq.~\ref{eq:Rabieff}, one immediately get the effective Rabi frequency $\Omega_{Rabi}^{eff}$. Its laser intensity dependency is plotted in Fig.~\ref{fig:split} as the orange broken line. The agreement between the measured photoelectron kinetic energy splitting and the effective Rabi frequency taken into account the effect of damping due to ionization is excellent, which shows clearly the effect of the ionization on the effective coupling strength. 




In conclusion, we have uncover a new source for tuning the effective Rabi frequency in the resonance multiphoton ionization of potassium atoms. The coupling strength between two resonantly coupled state are directly mapped out in the kinetic energy distribution of the photoelectrons. The coupling strength deviates significantly from the simple Rabi frequency as the laser intensity increases due to the non-negligible contribution of the ionization rate. Our results shows the possibility to control the interaction between laser field and the atom system with the cost of populations. This finding might offer new tuning nobs to the fields of manipulating atom with ultrafast laser pulses. For example, a second control laser can be used to drive the populations into some dark states where the effective Rabi frequency can be tuned. Moreover, such control can be implemented on a femtosecond or even attosecond time scale due to the rapid progress of the ultrafast laser techniques. We would also like to point out that identifying ionization channel based on the kinetic energy spectrum of the emitted electrons in strong field ionization process when the system has complex energy levels should receive extra cares. Not only the multiple resonances might be involved in the ionization and make the kinetic energy shift or split significantly, but also the detection procedure itself can induce considerable energy shifts.

\begin{acknowledgments}
We thank Prof. Songbin Zhang, Prof. Chuancun Shu and Prof. Difa Ye for very stimulating discussions. We acknowledge the support from National Key R\&D Program of China (No.~2019YFA0307701). D. Z. acknowledges the starting grant finical support from Jilin University. L. H. and P. M. acknowledge the support by the National Natural Science Foundation of China (No.~11704147 and No.~11904120).
\end{acknowledgments}


\bibliographystyle{iopart-num}
\bibliography{ref}

\end{document}